\documentclass[twocolumn,aps,superscriptaddress,showpacs, showkeys, nofootinbib,floatfix]{revtex4-1}
\usepackage{graphicx}
\usepackage[normalem]{ulem}  % \sout{old text} for strikeout
\usepackage[dvips]{color} % For blue in-text comments and additions

\renewcommand\sout{\bgroup \color{red} \ULdepth=-.5ex \ULset}

\begin{document}

\title{$\Omega_{ccc}$ Production in High Energy Nuclear Collisions }

\author{Hang He$^1$, Yunpeng Liu$^{2}$, and Pengfei Zhuang}

\address{Physics Department, Tsinghua University and Collaborative Innovation Center of Quantum Matter, Beijing 100084, China\\
         $^2$ Department of Physics, Tianjin University, Tianjin 300072, China}

\date{\today}

\begin{abstract}
We investigate the production of $\Omega_{ccc}$ baryon in high energy nuclear collisions via quark coalescence mechanism. The wave function of $\Omega_{ccc}$ is solved from the Schr\"odinger equation for the bound state of three charm quarks by using the hyperspherical method. The production cross section of $\Omega_{ccc}$ per binary collision in a central Pb+Pb collision at  $\sqrt{s_{NN}}=2.76$ TeV reaches 9 nb, which is at least two orders of magnitude larger than that in a p+p collision at the same energy. Therefore, it is most probable to discover  $\Omega_{ccc}$ in heavy ion collisions at LHC, and the observation will be a clear signature of the quark-gluon plasma formation.
\end{abstract}
\maketitle

From Quantum Chromodynamics (QCD) at finite temperature, it is widely accepted that there exists a deconfinement phase transition from hadron gas to a quark-gluon plasma (QGP) at a critical temperature $T_c \sim 155$ MeV~\cite{lattice1}. Such a phase transition is expected to be realized in heavy ion collisions at the Relativistic Heavy Ion Collider (RHIC) and the Large Hadron Collider (LHC). Since the fireball formed in a heavy ion collision expands rapidly, one cannot observe directly the QGP in the final state and needs probes to signal the QGP formation in the early stage of the fireball evolution. The quarkonium suppression is considered as such a sensitive probe~\cite{matsui}. The measured $J/\psi$ nuclear modification factor and especially the transverse momentum distributions at RHIC~\cite{star1,star2,phenix1,phenix2} and LHC~\cite{alice1,alice2,cms1} show a strong hot medium effect.

In this paper we investigate $\Omega_{ccc}$ production as an alternative probe of the QGP formation in heavy ion collisions at LHC. The existence of $\Omega_{ccc}$ baryon, the ground bound state of three charm quarks, is a direct result of the quark model. In p+p collisions at LHC energy, the $\Omega_{ccc}$ production is difficult, since it requires at least three pairs of charm quarks with small relative momenta in an event. In relativistic heavy ion collisions, however, there are plenty of off-diagonal charm quarks in the fireball, and the $\Omega_{ccc}$ production becomes much easier. The coalescence mechanism~\cite{fries1} has been successfully used to describe the light hadron production, especially the quark number scaling of the elliptic flow~\cite{molnar} and the enhancement of the baryon to meson ratio~\cite{fries2,hwa,greco1}. Taking into account this mechanism, the yield of $\Omega_{ccc}$ is proportional to the cube of the charm quark number, $N_{\Omega_{ccc}}\sim N_c^3$, at given temperature and volume of the fireball. For central Pb+Pb collisions at LHC energy, the $\Omega_{ccc}$ production becomes significant and may play an important role in the probe of QGP. The coalescence mechanism~\cite{greco2} or statistical emission~\cite{pbm,gorenstein,andronic} or regeneration~\cite{grandchamp,thews,zhu} for quarkonium production is widely discussed in heavy ion collisions and successfully explains the $J/\psi$ yield and momentum distributions. Recently it was suggested that $B_c$ mesons can be observed at RHIC and LHC due to the regeneration mechanism~\cite{schroedter,liu}.
The production of particles with double, triple and hidden charm in heavy ion collisions was also studied in the framework of a statistical coalescence model~\cite{kostyuk}, and the symmetries of the three-heavy-quark system was also investigated within the effective field theory framework of potential nonrelativistic QCD~\cite{brambilla}.

In coalescence models the change in the constituent distribution before and after the coalescence process is required to be small, namely the number of constituents involved in the coalescence must be small compared with the total constituent number of the system. In this sense, the coalescence mechanism is more suitable for the production of rare particles like $\Omega_{ccc}$. The coalescence probability in phase space, namely the Wigner function, is usually parameterized as a Gaussian distribution~\cite{greco2,chen} and the width is fixed by fitting the data in heavy ion collisions. For $\Omega_{ccc}$, there are currently no data, and an adjustable coalescence probability will lose the prediction power of the calculation. Fortunately, for charmed hadrons like $J/\psi$ and $\Omega_{ccc}$ we can calculate their wave function and in turn the Wigner function by solving the Schr\"odinger equation with the help of the lattice simulated heavy quark potential at finite temperature~\cite{petreczky}.

In the following we first solve the three-body Schr\"odinger equation via hyperspherical method to get the wave function in coordinate space and the Wigner function in phase space of $\Omega_{ccc}$, and then fix the coalescence hypersurface and derive the $\Omega_{ccc}$ momentum distribution via the coalescence mechanism. We will numerically calculate the $\Omega_{ccc}$ production in Pb+Pb collisions at LHC energy and summarize the results and physics in the end.

Since charm quarks are so heavy, we can employ the non-relativistic Schr\"odinger equation in the coordinate representation to describe the bound states of three charm quarks,
\begin{eqnarray}
\label{s1}
&& \hat H\Psi({\bf r}_1,{\bf r}_2,{\bf r}_3)=E_T\Psi({\bf r}_1,{\bf r}_2,{\bf r}_3),\nonumber\\
&& \hat H= \sum_{i=1}^3{\hat{\bf p}_i^2\over 2m_c} + V({\bf r}_1, {\bf r}_2, {\bf r}_3)
\end{eqnarray}
with charm quark mass $m_c$ and total energy $E_T$. As a usually used approximation~\cite{nielsen}, we neglect the three-body interaction and express the potential as a sum of pair interactions,
\begin{equation}
\label{sum}
V({\bf r}_1, {\bf r}_2, {\bf r}_3)=\sum_{i<j}V_{cc}({\bf r}_i,{\bf r}_j).
\end{equation}
According to the leading order QCD, the diquark potential is only one half of the quark-antiquark potential, $V_{cc}=V_{c\bar c}/2$. We assume that such a relation still holds in the case of strong coupling and take the Cornell potential
\begin{equation}
\label{vccbar}
V_{c\bar c}({\bf r}_i,{\bf r}_j)= -{\alpha\over |{\bf r}_{ij}|}+\sigma |{\bf r}_{ij}|,
\end{equation}
where ${\bf r}_{ij}={\bf r}_i-{\bf r}_j$ is the relative distance between the two quarks $i$ and $j$, and $\alpha=\pi/12$ and $\sigma=0.2$\ GeV$^2$ are coupling parameters of the potential which together with the charm quark mass $m_c=1.25$ GeV reproduce well the $J/\psi$ and $\Upsilon$ masses~\cite{satz} in vacuum. In hot and dense medium, the strength of the interaction between two quarks should decrease with temperature. However, from the lattice calculation~\cite{asakawa}, the $J/\psi$ spectral function is clearly broadened only at $T>T_c$. Therefore, we  still take the Cornell potential between a pair of charm quarks at the coalescence which happens at $T_c$.

It is hard to solve a three-body problem exactly, and one usually take some approximations to simplify the problem. One of the most popular and effective approaches is the hyperspherical method~\cite{nielsen}. Its main idea is to change a low dimensional three-body problem to a high dimensional one-body problem with the assumption of hyperspherical symmetry for the potential~\cite{narodetskii}. Since the potential (\ref{sum}) is only related to the relative coordinates ${\bf r}_i-{\bf r}_j$, the motion of the three-quark bound state can be factorized into the motion of the baryon and the relative motion among the quarks, by making the transformation between ${\bf r}_1, {\bf r}_2, {\bf r}_3$ and the baryon coordinate ${\bf R}$ and relative coordinates ${\bf r}_x, {\bf r}_y$,
$({\bf R}, {\bf r}_x, {\bf r}_y)=({\bf r}_1, {\bf r}_2, {\bf r}_3)M^T$ with the transformation matrix
\begin{equation}
\label{trans}
M=\left(\begin{array}{ccc}\frac{1}{3} & \frac{1}{3} & \frac{1}{3} \\\sqrt{\frac{1}{2}} & -\sqrt{\frac{1}{2}} & 0 \\\sqrt{\frac{1}{6}} & \sqrt{\frac{1}{6}} & -\sqrt{\frac{2}{3}}\end{array}\right).
\end{equation}
Then by rewriting ${\bf r}_x$ and ${\bf r}_y$ in terms of their azimuthal angles ${\theta_x, \phi_x}, {\theta_y, \phi_y}$ and the hyperradius $r=\sqrt{{\bf r}_x^2+{\bf r}_y^2}=\sqrt{\left({\bf r}_{12}^2+{\bf r}_{23}^2+{\bf r}_{31}^2\right)/3}$ and hyperpolar angle $\alpha=\text {arctan}(|{\bf r}_y|/|{\bf r}_x|)$, the volume element in hyper coordinates is represented as
\begin{equation}
\label{v}
d^3{\bf r}_xd^3{\bf r}_y = r^5 \sin^2\alpha\cos^2\alpha \sin\theta_x \sin\theta_y dr d\alpha d\theta_x d\phi_x d\theta_y d\phi_y
\end{equation}
and the kinetic energy in center of mass frame becomes
\begin{eqnarray}
\label{kin}
\hat T &=& {1\over 2m_c}\left(-{\partial^2\over\partial r^2}-{5\over r}{\partial\over\partial r}+{{\hat L}^2\over r^2}\right),\nonumber\\
{\hat L}^2 &=& -{\partial^2\over \partial\alpha^2}-4\text {cot}2\alpha{\partial\over\partial\alpha}+{{\hat l}_x^2\over \sin^2\alpha}+{{\hat l}_y^2\over \cos^2\alpha},
\end{eqnarray}
where $\hat L$ is the hyper angular momentum and $\hat l_x$ and $\hat l_y$ are the normal angular momenta.

Since the potential $V(|{\bf r}_i-{\bf r}_j|)$ depends on both the radius and the 5 angles, one can not directly separate the relative motion into a radial part and an angular part. The  approximation~\cite{narodetskii} we take here is to average the potential over all the angles,
\begin{equation}
\label{rv}
v(r)={8\over \pi}\int_0^{\pi/2}\sum_{i<j}V_{cc}\left(\sqrt 2 r\sin\alpha\right)\sin^2\alpha\cos^2\alpha d\alpha.
\end{equation}
With this homogeneous potential, the equation of relative motion can now be factorized into the radial equation (for the ground state with $L=0$)
\begin{equation}
\label{radial}
\left[{1\over 2m_c}\left(-{d^2\over dr^2}-{5\over r}{d\over dr}\right)+v(r)\right]\varphi(r)=E\varphi(r)
\end{equation}
and the angular equation
\begin{equation}
\label{angle}
\hat L^2 Y(\Omega)=L(L+4)Y(\Omega),
\end{equation}
where $\varphi(r)$ is the radial wave function, $Y(\Omega)$ the eigenstate of the hyper angular momentum operator $\hat L^2$ with $\Omega$ representing all the angle variables $\{\alpha,\theta_x,\phi_x,\theta_y,\phi_y\}$, $L$ the corresponding angular momentum number, and $E$ the relative energy.

\begin{figure}[!hbt]
\centering
\includegraphics[width=0.45\textwidth]{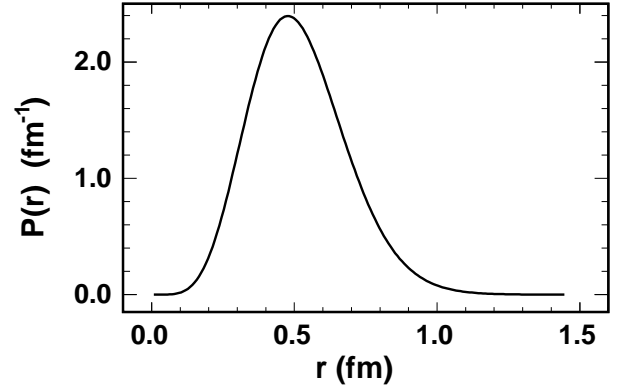}
\caption{The radial probability $P(r)$ to find the three charm quarks in the ground bound state in a hyper spherical shell of unit thickness at radius $r$. }
\label{fig1}
\end{figure}

The radial wave function $\varphi(r)$ is normalized as
\begin{equation}
\int_0^{\infty} P(r) dr = 1,
\label{eq_pr}
\end{equation}
where $P(r)=|\varphi(r)|^2 r^5$ is the probability to find the three charm quarks in the ground bound state in a hyper spherical shell of unit thickness at radius $r$. By solving the radial equation for $\Omega_{ccc}$, we obtain its mass $m_\Omega=4.7$ GeV and binding energy $\epsilon_\Omega = 900$ MeV. From the radial probability shown in Fig.\ \ref{fig1}, $\Omega_{ccc}$ is a tightly bound state of three charm quarks with average radius $r\sim 0.5$ fm which is almost the same as $J/\psi$.

We now construct the Wigner function in the center of mass frame of $\Omega_{ccc}$,
\begin{equation}
\label{wigner1}
W({\bf r},{\bf p})=\int d^6{\bf y} e^{-i{\bf p}\cdot{\bf y}}\psi\left({\bf r}+{{\bf y}\over 2}\right)\psi^*\left({\bf r}-{{\bf y}\over 2}\right),
\end{equation}
where ${\bf p}=({\bf p}_x,{\bf p}_y)$ is the 6D relative momentum corresponding to ${\bf r}=({\bf r}_x,{\bf r}_y)$, and the 3D relative momenta ${\bf p}_x, {\bf p}_y$ and the $\Omega_{ccc}$ momentum ${\bf P}$, corresponding to ${\bf r}_x, {\bf r}_y$ and ${\bf R}$, are associated with the three quark momenta ${\bf p}_1, {\bf p}_2, {\bf p}_3$ via the transformation $({\bf P},{\bf p}_x,{\bf p}_y)=({\bf p}_1,{\bf p}_2,{\bf p}_3)M^{-1}$. Using the above obtained relative wave function $\psi({\bf r})=\varphi(r)Y(\Omega)$ in the approximation of hyperspherical symmetry and taking the first axis of the vector ${\bf y}$ in the direction of ${\bf p}$ and the second axis on the plane constructed by ${\bf p}$ and ${\bf r}$, the Wigner function for the ground bound state $\Omega_{ccc}$ ($Y(\Omega)=\pi^{-3/2}$) is simplified as
\begin{eqnarray}
\label{wigner2}
&& W(r,p,\theta) = {1\over\pi^3}\int d^6{\bf y} e^{-ipy_1}\varphi\left(r_y^+\right)\varphi^*\left(r_y^-\right),\nonumber\\
&& r_y^\pm = r^2+{1\over 4}\sum_{i=1}^6y_i^2 \pm y_1r\cos\theta+y_2r\sin\theta.
\end{eqnarray}
Note that the vectors ${\bf r}$ and ${\bf p}$ in the Wigner function are correlated with each other through the angle $\theta$ between them. By integrating out the angle we obtain the probability to find the three charm quarks in the ground bound state in a hyperspherical shell in coordinate space at radius $r$ and in a hyperspherical shell in momentum space at radius $p$,
\begin{equation}
\label{pw}
{\cal P}(r,p) = {1\over 24\pi}r^5p^5\int_0^\pi W(r,p,\theta)\sin^4\theta d\theta
\end{equation}
which satisfies the normalization
\begin{equation}
\label{norm}
\int_0^\infty {\cal P}(r,p)dr dp=1.
\end{equation}

\begin{figure}[!hbt]
\centering
\includegraphics[width=0.50\textwidth]{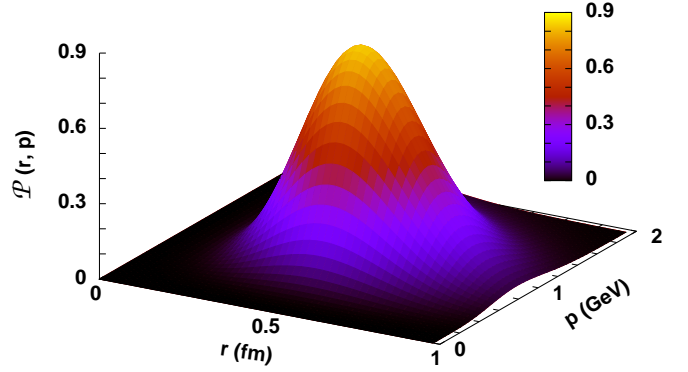}
\caption{The probability ${\cal P}(r,p)$ to find the three charm quarks in the ground bound state in a hyper spherical shell in coordinate space at radius $r$ and in a hyper spherical shell in momentum space at radius $p$. }
\label{fig2}
\end{figure}

Fig.\ \ref{fig2} shows the probability ${\cal P}(r,p)$. The most probable position in the phase space is located at $(r,p)\sim (0.5\ \text {fm}, 1\ \text {GeV})$, leading to $r\cdot p\approx 2.5$, which is near to the result $\sqrt{\langle r^2\rangle\langle p^2\rangle}=3$ from the uncertainty relation for a Gaussian Wigner function.

The observed momentum distribution of $\Omega_{ccc}$ via coalescence mechanism can be calculated from the Wigner function~\cite{fries3,scheibl},
\begin{eqnarray}
{d N\over d^3{\bf P}}&=& C \int {d^3{\bf R}\over (2\pi)^3}\int {d^3{\bf r}_x d^3{\bf r}_y d^3{\bf p}_x d^3{\bf p}_y\over (2\pi)^6}\nonumber\\
&\times& F({\bf r}_1,{\bf r}_2,{\bf r}_3,{\bf p}_1,{\bf p}_2,{\bf p}_3)W({\bf r}_x,{\bf r}_y,{\bf p}_x,{\bf p}_y),
\label{co1}
\end{eqnarray}
where $F$ is the distribution function of the three charm quarks in phase space, $C$ the factor to count the intrinsic symmetry. For $\Omega_{ccc}$, it is a color singlet and carries spin $3/2$. Since there is only one color singlet state in the $3\times3\times3=27$ possible color states and $4$ spin $3/2$ states in the $2\times 2\times 2=8$ possible spin states, we obtain $C=1/27\times 4/8=1/54$.

In relativistic heavy ion collisions, the hadronization of the parton system happens on the hypersurface of confinement phase transition. The 4D coordinates $R_{\mu}=(t,{\bf R})$ on the hypersurface is constrained by the hydronization condition,
\begin{equation}
T(R_\mu)= T_c
\label{tc}
\end{equation}
which leads to $t=t(T_c,{\bf R})$, where $T_c$ is the critical temperature of the confinement phase transition, and the local temperature $T(R_\mu)$ and fluid velocity $u_\mu(R_\mu)$ (which will be used in the charm quark distribution) are determined by hydrodynamic equations
\begin{equation}
\partial_{\mu}T^{\mu\nu}= 0
\end{equation}
with $T^{\mu\nu}=(\epsilon+p)u^{\mu}u^{\nu}-g^{\mu\nu}p$ being the energy momentum tensor and $\epsilon$ and $p$ the energy density and pressure. To close the hydrodynamical equations one needs to know the equation of state of the medium. We follow Ref.\ \cite{sollfrank} where the
deconfined phase at high temperature is an ideal gas of gluons and massless
$u$ and $d$ quarks plus 150 MeV massed $s$ quarks, and the
hadron phase at low temperature is an ideal gas of all known
hadrons and resonances with mass up to 2 GeV~\cite{pdg}. There is
a first order phase transition between these two phases with the
critical temperature $T_c=165$ MeV. For the initialization of the hot medium, we take the same treatment as in Ref.~\cite{hirano}. The maximum temperature of the medium at the starting time $\tau_0=0.6$ fm/c is $T_0=484$ MeV for central 2.76 TeV Pb+Pb collisions at LHC.

Changing the volume integral $d^3{\bf R}$ to the covariant integral on the hypersurface $\Sigma$, the $\Omega_{ccc}$ distribution is rewritten as
\begin{eqnarray}
{dN\over d^2{\bf P}_Td\eta} &=& C\int_\Sigma{P^\mu d\sigma_\mu(R)\over (2\pi)^3}\int {d^4r_x d^4r_y d^4p_x d^4p_y\over (2\pi)^6}\nonumber\\
&\times&F(\tilde r_1,\tilde r_2,\tilde r_3,\tilde p_1,\tilde p_2,\tilde p_3)W(r_x,r_y,p_x,p_y),\nonumber\\
\label{co2}
\end{eqnarray}
where ${\bf P}_T, \eta$ and $P^0=\sqrt{{\bf P}^2+m_\Omega^2}$ are, respectively, the $\Omega_{ccc}$ transverse momentum, rapidity and energy. Remember that the Wigner function obtained above is derived in the center of mass frame of $\Omega_{ccc}$ and the $\Omega_{ccc}$ moves with 4-velocity $v^\mu=P^\mu/m_\Omega$ in the laboratory frame, the coordinates ${\bf r}_1, {\bf r}_2, {\bf r}_3$ or ${\bf r}_x, {\bf r}_y$ in the parton distribution function $F$ should be replaced by $\tilde r_x^\mu=L^\mu_{\ \nu} r_x^\nu, \tilde r_y^\mu=L^\mu_{\ \nu} r_y^\nu$ with the boost matrix elements $L^0_{\ 0}=v^0$, $L^0_{\ i}=L^i_{\ 0}=v^i$, $L^i_{\ j}=\delta_{i,j}+\xi v^iv^j$, and $\xi=1/(1+v^0)$. Since in the center of mass frame of the three charm quarks the coalescence happens at the same time, there is $r_x^0=r_y^0=0$. Similarly, the momenta ${\bf p}_i\ (i=1,2,3)$ in the charm quark distribution $F$ are replaced by $\tilde p_i^\mu=L^\mu_{\ \nu} p_i^\nu$ with $p^0=\sqrt{{\bf p}^2_i+m_c^2}$.

We now consider the integral element $d\sigma_\mu(R)$ over the coalescence hypersurface $\Sigma$. In the framework of Bjorken hydrodynamics~\cite{bjorken}, we take the rapidity $\eta=(1/2)\ln[(t+R_z)/(t-R_z)]$, the transverse radius $R_T=\sqrt{R_x^2+R_y^2}$ and the azimuth angle $\phi=\arctan (R_y/R_x)$ as independent variables instead of ${\bf R}$ and regard the proper time $\tau=\sqrt{t^2-R_z^2}$ as a function of $\eta, R_T, \phi$ through the coalescence condition (\ref{tc}), the hypersurface element can be expressed as
\begin{eqnarray}
\label{dsigma}
d\sigma_0 &=& \left(R_T{\partial\tau\over \partial\eta}\sinh\eta+R_T\tau\cosh\eta\right)dR_Td\phi d\eta,\nonumber\\
d\sigma_1 &=& \left(\tau{\partial\tau\over \partial\phi}\sin\phi-R_T\tau{\partial\tau\over\partial R_T}\cos\phi\right)dR_Td\phi d\eta,\nonumber\\
d\sigma_2 &=& -\left(\tau{\partial\tau\over \partial\phi}\cos\phi+R_T\tau{\partial\tau\over\partial R_T}\sin\phi\right)dR_Td\phi d\eta,\nonumber\\
d\sigma_3 &=& -\left(R_T{\partial\tau\over \partial\eta}\cosh\eta+R_T\tau\sinh\eta\right)dR_Td\phi d\eta.
\end{eqnarray}

The three quark distribution $F(r_1,r_2,r_3,p_1,p_2,p_3)$ can be factorized as
\begin{equation}
F(r_1,r_2,r_3,p_1,p_2,p_3)= Sf(r_1,p_1)f(r_2,p_2)f(r_3,p_3),
\end{equation}
where $S$ counts the symmetry of the same specie of quarks. For $\Omega_{ccc}$ we simply take $S=1/3!=1/6$, since the number of charm quarks in an event is much larger than 3 at LHC energy. The single charm quark distribution $f$ is in principle between the pQCD and equilibrium distributions. From the experimental
data at LHC~\cite{lhc1,lhc2}, the observed large quench factor and elliptic flow for charmed mesons indicate that the charm quarks interact strongly with the medium.
Therefore, one can take, as a good approximation, a kinetically thermalized phase space
distribution for charm quarks,
\begin{equation}
   f(r_i,p_i)= \rho(r_i){N(r_i)\over e^{u^\mu(r_i) p^i_\mu/T(r_i)}+1},
\end{equation}
where the local temperature $T(r_i)$ and fluid 4-velocity $u_\mu(r_i)$ of the medium are determined by the hydrodynamics, and
\begin{equation}
   N(r_i)=\left[\int {d^3{\bf p}\over (2\pi)^3}\frac{1}{e^{u^{\mu}(r_i)p_{\mu}/T(r_i)}+1}\right]^{-1}
\end{equation}
is the normalization factor. The number density $\rho(r_i)$ is controlled by the charm conservation equation
\begin{equation}
   \partial_\mu\left[\rho(r_i)u^\mu(r_i)\right]= 0.
\end{equation}
The charm quark number density at initial time $\tau_0=0.6$\ fm/c is fixed by the colliding energy and nuclear geometry,
\begin{equation}
\rho(\tau_0, {\bf x}_T, \eta)={T_A({\bf x}_T)T_B({\bf x}_T-{\bf b})\cosh\eta\over \tau_0}{d\sigma^{c\overline c}_{pp}\over d\eta},
\end{equation}
where $T_A$ and $T_B$ are the thickness functions of the lead nuclei with nuclear matter density following the Woods-Saxon distribution, $d\sigma^{c\overline c}_{pp}/d\eta$ is the rapidity distribution of charm quark cross section in $p+p$ collisions, and ${\bf b}$ is the impact parameter.

\begin{figure}[!hbt]
\centering
\includegraphics[width=0.50\textwidth]{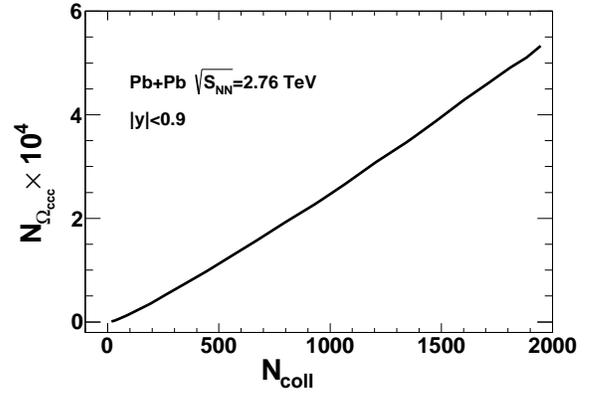}
\caption{The $\Omega_{ccc}$ yield as a function of the number of binary collisions $N_{coll}$ in Pb+Pb collisions at middle rapidity and colliding energy $\sqrt{s_{NN}}=2.76$\ TeV.
}
\label{fig3}
\end{figure}

We now apply the above coalescence approach to $\Omega_{ccc}$ production in relativistic heavy ion collisions. The yield at middle rapidity in Pb+Pb collisions at colliding energy $\sqrt {s_{NN}}=2.76$ TeV is shown in Fig.\ \ref{fig3} as a function of the number of binary collisions $N_{coll}$. The charm production cross section is taken as $d\sigma_{c\bar c}/d\eta= 0.7	$\ mb~\cite{alice3}. The yield increases almost linearly with $N_{coll}$ and reaches $5\times 10^{-4}$ in the most central collisions. If we consider a homogeneous fireball with volume $V$ at the coalescence time, the yield of $\Omega_{ccc}$ can be estimated as
\begin{equation}
   N_\Omega \sim {N_c^3\over V^2},
\end{equation}
where $N_c$ is the charm quark yield. Supposing both $N_c$ and $V$ are proportional to $N_{coll}$, the yield of $\Omega_{ccc}$ is then proportional to $N_{coll}$, which approximately explains the linear increase in Fig.\ \ref{fig3}. From the $\Omega_{ccc}$ yield we can define an effective cross section per binary collision,
\begin{equation}
\label{eff}
\sigma_\Omega \equiv {N_\Omega\over N_{coll}\Delta \eta}\sigma_{pp}.
\end{equation}
With the inelastic proton cross section $\sigma_{pp}=62$\ mb, the rapidity range $\Delta \eta=1.8$, and $N_{coll}=2000$ for the most central collision, we have $\sigma_\Omega= 9$\ nb, which is much larger than the cross section 0.06-0.13\  nb at 7 TeV and 0.1-0.2\ nb at 14 TeV in p+p collisions at mid rapidity $|\eta| < 2.5$~\cite{chen}. It is necessary to point out that the tightly bound states of heavy quarks are in principle continuously produced in the medium above $T_c$ and suffer from dissociation due to the interaction with the medium~\cite{yan}. Therefore, the above obtained $\Omega_{ccc}$ yield from the sudden coalescence approach without considering dissociation is more like the upper limit of the production.

\begin{figure}[!hbt]
\centering
\includegraphics[width=0.50\textwidth]{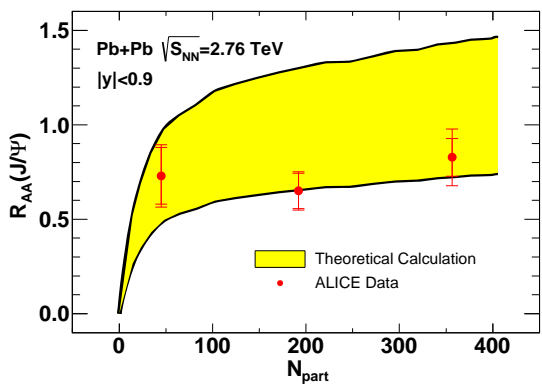}
\caption{The $J/\psi$ nuclear modification factor $R_{AA}$ at middle rapidity as a function of the number of participant $N_{part}$ in Pb+Pb collisions at $\sqrt{s_{NN}} = 2.76$ TeV. The theoretical band is due to the uncertainty in the charm cross section, the upper and lower limit of the band correspond to $d\sigma/d\eta = 0.7$ and $0.5$ mb. The data are from the ALICE collaboration~\cite{alice4}.}
\label{fig4}
\end{figure}

We also calculated the wave function and in turn the Wigner function for $J/\psi$ where the interaction between the $c$ and $\bar c$ is exactly the Cornell potential. The calculated nuclear modification factor $R_{AA}=N_{AA}/(N_{pp}N_{coll})$ for $J/\psi$ in Pb+Pb collisions at LHC energy is shown in Fig.~\ref{fig4} as a function of the number of participant nucleons $N_{part}$, where $N_{AA}$ and $N_{pp}$ are, respectively, the $J/\psi$ yield in Pb+Pb and p+p collisions. The model calculation with charm cross section $d\sigma/d\eta =0.7$ mb, corresponding to the upper limit of the theoretical band, is clearly overestimated, in comparison with the experimental data~\cite{alice4}. This is probably due to the lack of $J/\psi$ dissociation in the hot medium and the large charm cross section. Note that the $J/\psi$ production in heavy ion collisions is more complicated than $\Omega_{ccc}$. $J/\psi$s can be produced via both initial p+p collisions and later coalescence, while the coalescence is the only way for $\Omega_{ccc}$ production at LHC energy. A good description of the experimental $J/\psi$ data needs $d\sigma/d\eta=0.5$ mb in our calculation, see the lower limit of the theoretical band in Fig.\ \ref{fig4}.

\begin{figure}[!hbt]
\centering
\includegraphics[width=0.50\textwidth]{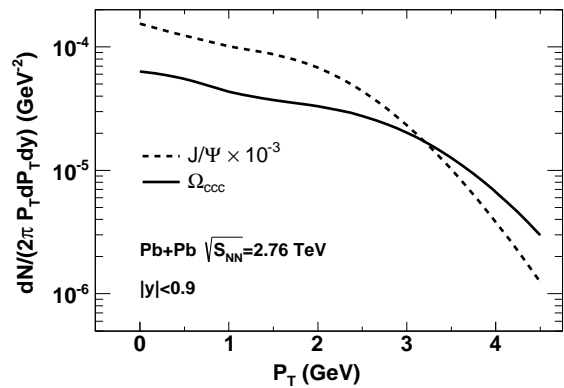}
\caption{The transverse momentum distribution of $\Omega_{ccc}$ (solid line) and $J/\psi$ (dashed line) at middle rapidity in central Pb+Pb collisions at $\sqrt{s_{NN}}=2.76$ TeV. The calculation for $J/\psi$ is scaled by a factor of $10^{-3}$.}
\label{fig5}
\end{figure}

In heavy ion collisions, transverse motion is developed during the dynamical evolution of the system. The
microscopically high particle density and multiple scatterings are essential for the finally observed transverse
momentum distributions. The distributions are therefore sensitive to the medium properties, like the equation
of state. In order to understand the $\Omega_{ccc}$ production mechanism and extract the
properties of the medium, we calculated the transverse momentum distributions of $\Omega_{ccc}$ and $J/\psi$,  shown in Fig.~\ref{fig5} with the assumption of thermalized charm quark distribution. In both cases the distribution drops down  monotonously with transverse momentum $p_T$. For $\Omega_{ccc}$ it is about one order of magnitude smaller at $p_T = 4$ GeV than that at $p_T=0$. As a characteristic of the coalescence mechanism~\cite{fries2,hwa,greco1}, the decreasing of $J/\psi$ becomes faster than $\Omega_{ccc}$ at high $p_T$.

In summary, we investigated the production of $\Omega_{ccc}$ baryon via coalescence mechanism in relativistic heavy ion collisions. We solved the Schr\"odinger equation for the ground bound state of three charm quarks by the hyperspherical method and derived the radial wave function of $\Omega_{ccc}$. With the obtained Wigner function as the coalescence probability and thermalized charm quark distribution, we calculated the $\Omega_{ccc}$ yield and transverse momentum spectrum in Pb+Pb collisions at colliding energy $\sqrt{s_{NN}}=2.76$ TeV. The obtained  production cross section per binary collision is at least two orders of magnitude larger than that in a p+p collision at LHC energy. Therefore, from the weak decay modes like the non-leptonic decay channel $\Omega_{ccc}\to\Omega_{sss}+3\pi^+$~\cite{chen}, it becomes most probable to observe $\Omega_{ccc}$ in heavy ion collisions at LHC, and its observation is a clear signature of the quark-gluon plasma formation.

\vspace{0.2cm}
\appendix {\bf Acknowledgement:}
We thank Liewen Chen for helpful discussions in the beginning of this work. The work is supported by the NSFC and MOST grant Nos. 11335005, 2013CB922000 and 2014CB845400.


\begin{thebibliography}{10}
\bibitem{lattice1} A.~Bazavov et al. (HotQCD Collaboration), Phys. Rev. {\bf D85}, 054503(2012).
\bibitem{matsui} T.~Matsui and H.~Satz, Phys. Lett. {\bf B178}, 417(1986).
\bibitem{star1} L.~Adamczyk et al. (STAR Collaboration), Phys. Lett. {\bf B722}, 55(2013).
\bibitem{star2} L.~Adamczyk et al. (STAR Collaboration), Phys. Rev. Lett. {\bf 111}, 052301(2013).
\bibitem{phenix1} A.~Adare et al. (PHENIX Collaboration), Phys. Rev. {\bf C84}, 054912(2011).
\bibitem{phenix2} A.~Adare et al. (PHENIX Collaboration), Phys. Rev. Lett. {\bf 98}, 232301(2007).
\bibitem{alice1} E.~Abbas et al. (ALICE Collaboration), Phys. Rev. Lett. {\bf 111}, 162301(2013).
\bibitem{alice2} B.~Abelev et al. (ALICE Collaboration) Phys. Lett. {\bf B743}, 314(2014).
\bibitem{cms1} S.~Chatrchyan et al. (CMS Collaboration), JHEP {\bf 05}, 063(2012).
\bibitem{fries1} R.~Fries, V.~Greco, and P.~Sorensen, Annual Review of Nuclear and Particle Science {\bf 58}, 177(2008).
\bibitem{molnar} D.~Molnar and S.~A.~Voloshin, Phys. Rev. Lett. {\bf 91}, 092301(2003).
\bibitem{hwa} R.~C.~Hwa and C.~B.~Yang, Phys. Rev. {\bf C67}, 034902(2003).
\bibitem{fries2} R.~J.~Fries, B.~Muller, C.~Nonaka, and S.~A.~Bass, Phys. Rev. Lett. {\bf 90}, 202303(2003).
\bibitem{greco1} V.~Greco, C.~M.~Ko, and P.~Levai, Phys. Rev. Lett. {\bf 90}, 202302(2003).
\bibitem{greco2} V.~Greco, C.~M.~Ko, and R.~Rapp, Phys. Lett. {\bf B595}, 202(2004).
\bibitem{pbm} P.~Braun-Munzinger and J.~Stachel, Phys. Lett. {\bf B490}, 196(2000).
\bibitem{gorenstein} M.~I.~Gorenstein, A.~Kostyuk, H.~Stoecker, and W.~Greiner, Phys. Lett. {\bf B509}, 277(2001).
\bibitem{andronic} A.~Andronic, P.~Braun-Munzinger, K.~Redlich, and J.~Stachel, Phys. Lett. {\bf B571}, 36(2003).
\bibitem{grandchamp} L.~Grandchamp and R.~Rapp, Nucl. Phys. {\bf A709}, 415(2002).
\bibitem{thews} R.~L.~Thews, M.~Schroedter, and J.~Rafelski, Phys. Rev. {\bf C63}, 054905(2001).
\bibitem{zhu} X.~Zhu, P.~Zhuang, and N.~Xu, Phys. Lett. {\bf B607}, 107(2005).
\bibitem{schroedter} M.~Schroedter, R.~L.~Thews, and J.~Rafelski, Phys. Rev. {\bf C62}, 024905(2000).
\bibitem{liu} Y.~Liu, C.~Greiner, and A.~Kostyuk, Phys. Rev. {\bf C87}, 014910(2013).
\bibitem{kostyuk} A.~Kostyuk, arXiv:nucl-th/0502005.
\bibitem{brambilla} N.~Brambilla, F.~Karbstein, and A.~vairo, Phys. Rev. {\bf D87}, 074014(2013).
\bibitem{chen} L.~W.~Chen and C.~M.~Ko, Phys. Rev. {\bf C73}, 044903(2006).
\bibitem{petreczky} P.~Petreczky, J. Phys. {\bf G37}, 094009(2010).
\bibitem{nielsen} E.~Nielsen, D.~fedorov, A.~Jensen, and E.~garrido, Phys. Rep. {\bf 347}, 373(2001).
\bibitem{satz} H.~Satz, J. Phys. {\bf G32}, R25(2006).
\bibitem{asakawa} M.~Asakawa and T.~Hatsuda, Phys. Rev. Lett. {\bf 92}, 012001(2004).
\bibitem{narodetskii} I.~Narodetskii, Y.~Simonov, and A.~Veselov, JETP Letters {\bf 90}, 232(2009).
\bibitem{fries3} R.~J.~Fries, B.~Muller, C.~Nonaka, and S.~A.~Bass, Phys. Rev. {\bf C68}, 044902(2003).
\bibitem{scheibl} R.~Scheibl and U.~W.~Heinz, Phys. Rev. {\bf C59}, 1585(1999).
\bibitem{sollfrank} J.~Sollfrank et al., Phys. Rev. {\bf C55}, 392(1997).
\bibitem{pdg} K.~Hagiwara et al. (Particle Data Group), Phys. Rev. {\bf D66}, 010001(2002).
\bibitem{hirano} T.~Hirano, P.~Huovinen, and Y.~Nara, Phys. Rev. {\bf C83}, 021902(2011).
\bibitem{bjorken} J.~D.~Bjorken, Phys. Rev. {\bf D27}, 140(1983).
\bibitem{lhc1} B.~Abelev et al. (ALICE Collaboration), JHEP {\bf 09}, 112(2012).
\bibitem{lhc2} B.~Abelev et al. (ALICE Collaboration), Phys. Rev. Lett. {\bf 111}, 102301(2013).
\bibitem{alice3} B.~Abelev et al.(ALICE Collaboration), JHEP {\bf 07}, 191(2012).
\bibitem{chen} Y.~Chen and S.~Wu, JHEP {\bf 08}, 144(2011).
\bibitem{yan} L.~Yan, P.~Zhuang, and N.~Xu, Phys. Rev. Lett. {\bf 97}, 232301(2006).
\bibitem{alice4} Pereira Da Costa Hugo et al. (ALICE Collaboration), arXiv:1110.1035, AIP Conf. Proc. {\bf 1441}, 859(2012).
\end{thebibliography}
\end{document}